\documentclass[a4paper]{amsart}
\pdfoutput=1 
\usepackage[utf8]{inputenc}
\usepackage[T1]{fontenc}
\usepackage{lmodern}

\usepackage{tikz}
\usepackage[ruled]{algorithm2e}

\usepackage{microtype}

\usepackage[backend=biber,alldates=edtf]{biblatex}
\usepackage{placeins}

\title{Estimation of Climbing Route Difficulty using Whole-History Rating}
\thanks{The author thanks Simon Dale and theCrag Pty Ltd.\@ (\url{https://www.thecrag.com}) for providing their database of ascents for analysis}
\author{Dean Scarff}

\usepackage{hyperref}
\hypersetup{pdftitle={Estimation of Climbing Route Difficulty using Whole-History Rating},pdfauthor={Dean Scarff}}

\newcommand{\theCrag}{\emph{\href{https://www.thecrag.com}{theCrag}}}

\begin{document}
\begin{abstract}
  Existing grading systems for rock climbing routes assign a difficulty grade to a route based on the opinions of a few people.  An objective approach to estimating route difficulty on an interval scale was obtained by adapting the Whole-History Rating (WHR) system to rock climbing.  WHR's model was fitted to a database of 236,095 ascents recorded by users on an established climbing website.  73\% of the ascents used in the dataset were classified as successful.  Predictions were on average 85\% accurate with 10-fold cross-validation.  The results suggest that an empirical rating system is accurate at assessing route difficulty and is viable for revising conventional route grades.
\end{abstract}

\maketitle

\section{Introduction}

Conventional systems for grading the difficulty of routes in outdoor rock climbing do not have a formal statistical basis.  The accepted grade for a route is determined based on the opinion of a few people; primarily the climbers who first ascended the route.  Grading systems around the world rely on subjective factors \cite{draper}.  John Ewbank, the originator of the de facto standard for grading free-climbing routes in Australia, New Zealand and South Africa, describes the qualitative factors that affect grades in his guidebook:
\begin{quotation}
  Grading takes the following into consideration.  Technical difficulty, exposure, length, quality of rock, protection and other smaller factors. \cite[11]{ewbank1967}
\end{quotation}

These grading systems have several deficiencies.  The subjectivity and influence of different raters for each route is a source of inconsistency: indeed, climbing slang has terms to describe routes that are relatively difficult (`sandbagged') or easy (`soft') compared to their claimed grade.  While Ewbank grades use an integer scale, there is no formal interpretation of the numerical difference or ratio between grades.  Subjective grades also lack refutability: they are a matter of opinion, not evidence.

Instead, it's desirable to rate route difficulty objectively, with measurable accuracy and an interval scale.  To this end, the applicability of the Whole-History Rating System (WHR) \cite{whr} to climbing was evaluated.  WHR, which uses a similar model to the Elo rating system \cite{elo}, is used for estimating the time-varying strength of players in two-player games.  This paper presents a method of adapting the WHR model to climbing and the results of fitting that model to a database of ascents.

The remainder of the paper is structured as follows.  Section~\ref{sec:method} describes the model, algorithm, dataset and model evaluation procedure.  Section~\ref{sec:results} explains key results, and Section~\ref{sec:discussion} discusses limitations of the model and opportunities for future research.

\section{Method}
\label{sec:method}

\subsection{Model}

The WHR model for climbing is largely the same as the dynamic Bradley-Terry model originally described for WHR \cite[\S2]{whr}.  This section summarizes the WHR model, and highlights how it can be adapted for rock climbing.

Climbers and routes are equivalent to players in WHR.  Route ratings are assumed to remain constant over time.  Successful ascents (where the climber completes the route without falling or resting on equipment) are classified as a win for the climber.

In the WHR model, players have a rating, $r$.  Let $A$ be a Bernoulli trial corresponding to the outcome of a given ascent.  For climbing, the probability of a successful ascent by a climber with rating $r_i$ on a route with rating $r_j$ is given by the Bradley-Terry model:
\[
  P(A=1) = \frac{\exp(r_i)}{\exp(r_i) + \exp(r_j)}.
\]

The prior distribution for climber and route ratings is normally distributed.  This is a departure from the original WHR model, which uses a prior for each player's initial rating equivalent to one virtual win and one virtual loss against a player of rating 0.  The normal distribution allows for intuitive and independent parameterization of the priors for climbers and routes.  For routes,
\[
  r_i \sim \mathcal{N}\mathopen{}\left( \mu_i, \sigma_R^2 \right),
\]
and for the initial rating of a climber,
\[
  r_i(t_1) \sim \mathcal{N}\mathopen{}\left( 0, \sigma_C^2 \right).
\]
Using a mean of zero for climbers promotes numerical stability and identifiability.

The model defines a weight parameter $b$ that informs the priors on route ratings using the conventional grade $g$.  With a reference grade $g_0$, the mean parameter of route $i$'s prior is $\mu_i = b (g_i - g_0)$.
This particular functional form was chosen after observing a linear relationship between ratings and conventional grades.

A Wiener process with variance $w^2$ models the rating of a climber over time.  Hence if the rating of a given climber at time $t$ is $r(t)$,
\[
  r(t_2) - r(t_1) \sim \mathcal{N}\mathopen{}\left( 0, |t_2 - t_1|w^2 \right).
\]

For a set of ascents $\mathbf{A}$ and ratings $\mathbf{r}$, the model defines the posterior probability density function
\[
  f(\mathbf{r}) = p(\mathbf{r} | \mathbf{A}) =
  \frac{P(\mathbf{A} | \mathbf{r}) p(\mathbf{r})}
  {p(\mathbf{A})}.
\]

Here $P(\mathbf{A} | \mathbf{r})$ is the likelihood given by the Bradley-Terry model, and $p(\mathbf{r})$ is the probability density function given by the normal and Wiener priors.  $p(\mathbf{A})$ is a normalizing constant.

\subsection{Algorithm}

WHR was implemented largely as originally described \cite[\S3]{whr}.  This section summarizes the WHR algorithm and highlights implementation choices.

As input, the algorithm takes a set of ascent tuples consisting of the ascent outcome, climber, route, and a time period.  The algorithm produces maximum a posteriori estimates of climber and route ratings.  A climber may have multiple ratings, corresponding to each discrete time period where that climber has ascents.  The route and climber ratings can be used to generate an estimate of the probability a climber at a given time will be able to climb a particular route successfully.

\begin{algorithm}
  \caption{Whole-History Rating for climbing}
  \label{alg:whr}
  \KwIn{set of ascents $\mathbf{A}$}
  \KwOut{vector of route and climber ratings $\mathbf{r}$}
  initialize $\mathbf{r}$ from priors\;
  \While{$P(\mathbf{A} | \mathbf{r})$ has not converged}{
    \ForEach{climber rating $i$}{
      $r_i \gets r_i - \partial / \partial r_i \log f(r_i) 
      / \left( \partial^2 / \partial r_i^2 \log f(r_i) \right)$\;
    }
    \ForEach{route $i$}{
      $r_i \gets r_i - \partial / \partial r_i \log f(r_i) 
      / \left( \partial^2 / \partial r_i^2 \log f(r_i) \right)$\;
    }
  }
\end{algorithm}

The high-level algorithm (Algorithm~\ref{alg:whr}) works by iteratively adjusting a vector of all ratings $\mathbf{r}$ to maximize the likelihood function $f(\mathbf{r})$.  The normalizing constant $p(\mathbf{A})$ from the posterior probability density function is not evaluated.  The ratings adjustments are calculated using Newton's method.  WHR decomposes the Hessian to efficiently compute the derivative terms for the Wiener process \cite[Appendix~B]{whr}.  The algorithm terminates when the marginal log-likelihood from the Bradley-Terry model has not changed by more than 1 unit in the last 8 iterations.  The resulting vector $\mathbf{r}$ is the maximum a posteriori estimate of the ratings.

Route ratings are adjusted separately from the climber ratings, unlike in the original WHR algorithm where all players are adjusted in one loop.  This has two advantages.  Firstly, in a given iteration of the outer loop, the adjustments to the route ratings will be based on the latest estimate of all climber ratings, so it removes ordering as a source of non-determinism.  Secondly, evaluation of the Wiener prior can be skipped for routes.

The WHR algorithm was implemented\footnote{full source code available at \url{https://github.com/p00ya/climbing_ratings}} with the \textit{Python~3} programming language \cite{python}, the \textit{numpy} software package \cite{numpy}, and the \textit{Cython} programming language \cite{cython}.  The implementation uses 64-bit (double) precision floating point numbers, with the exception of the exponential terms used in the evaluation of the Bradley-Terry model, which use 80-bit (extended) precision floating point numbers.  It was not necessary to subtract small values from the diagonal of the Hessian matrix for numerical stability.  The space complexity is linear.  Each iteration of the outer loop has linear time complexity.  Computations were vectorized where practical.

\subsection{Data}

A database of ascents was obtained from \theCrag{}, a website that states its mission is ``to build an enduring resource of the world's climbing information, to facilitate sustainable climbing and to support a thriving community'' \cite{thecrag}.

Climbers can register and record their ascents on the website.  Only ascents from climbers who marked their logbook as publicly accessible were used.  Ascent records include a `tick type' indicating the success of the ascent \cite{thecrag:ticktypes}, the date of the ascent, the route climbed, and the conventional grade of the route (climber-rated, defaulting to the route's `assigned' grade).  Climbers choose which routes they climb, which ascents they record, and what tick type and date they associate with recorded ascents.

Only routes in Australia with a `natural' (outdoor) setting and a `sport' gear style \cite{thecrag:gearstyles} were selected.  The model's one-dimensional difficulty rating is inconsistent with the different skill sets required for different gear styles.  Routes were limited to Australia because of the consistent use of Ewbank grades and the density of the ascent data available there.

The ascents were preprocessed.  Tick types were reclassified as successful, unsuccessful, or ambiguous.  Ambiguous ascents were removed.  Ascents on routes that were not graded in the Australian (Ewbank) system were also filtered out.  The median Ewbank grade from the ascents of a route was used as the route's conventional grade $g$.

Ascent dates were quantized to a one-week resolution, as opposed to the one-day resolution used by WHR.  One week was chosen to avoid overfitting.  If a climber attempts a route over several days until they succeed, it's desirable for the probability of success to be modeled on that set of attempts as a whole, without attributing the final success to an intrinsic change in climbing skill.

Routes with less than two ascents were removed, because the model will be dominated by its priors for such routes, limiting the useful information gained.  Climbers with only successful ascents were also removed, because this indicates a reporting bias.

The Australian dataset consists of 236,095 ascents by 3,000 climbers, over 8,917 routes (332,956 ascents by 5,778 climbers, over 11,351 routes before preprocessing).  An `NSW' subset was used for training, using only ascents from New South Wales, consisting of 135,255 ascents by 1,958 climbers, over 5,293 routes (196,602 ascents by 3,719 climbers, over 6,869 routes before preprocessing).

\subsection{Model Training \& Evaluation}

Suitable values of the hyperparameters $\sigma_R^2$, $\sigma_C^2$, $b$ and $w^2$ were chosen at the discretion of the author based on exploration of the NSW dataset.  Accuracy was defined using a binary classifier that predicts ascents as successful where the Bradley-Terry model predicts $P(A = 1) > 0.5$, as for the original WHR experiment \cite[\S4]{whr}.  Adaptive, stratified 10-fold cross-validation, as implemented by the \textit{caret} software package \cite{caret}, was used to measure performance with different hyperparameters.

The chosen hyperparameters were then used to fit the model to the Australian dataset.  Stratified, 10-fold cross-validation with 3 repeats was used for validation, along with diagnostics comparing the residuals to estimated ratings, and estimated ratings to conventional Ewbank grades.

A naïve model predicting $P(A = 1) = \bar{A}$ was used as a performance baseline, due to the imbalanced success rate $\bar{A}$.  Precision and accuracy have a baseline of $\bar{A}$, and the balanced accuracy baseline is 50\%.  The baseline log loss is $(\bar{A} - 1) \log (1 - \bar{A}) - \bar{A} \log (\bar{A})$.

\section{Results}
\label{sec:results}

This section describes the chosen hyperparameters, and then goes on to describe the results after fitting the model to the Australian dataset.

\begin{table}
  \caption{Hyperparameter values chosen after experiments with the NSW dataset}
  \label{tab:hyperparam}
  \begin{tabular}{lll}
    \hline
    Symbol & Value & Definition \\ \hline
    $\sigma_C^2$ & 1 & initial climber rating variance \\
    $\sigma_R^2$ & 4 & route rating variance \\
    $w^2$ & $1/52$ & climber rating variance per week \\
    $g_0$ & 22 & reference Ewbank grade \\
    $b$ & 0.4 & conventional grade prior weight \\
    \hline
  \end{tabular}
\end{table}

\subsection{Hyperparameters}

Table~\ref{tab:hyperparam} summarises the chosen hyperparameter values after evaluating the model performance on the NSW dataset.

Grade 22 was chosen as the reference grade $g_0$ because it was the most common route grade.  This choice aligns the mode of the ratings prior distribution with the mode of the Ewbank grades, at $r = 0$.  Concentrating the mass of the ratings distribution near zero promotes numerical stability.

The variance of the climber ratings process $w^2 = 1/52$ can be interpreted as a standard deviation in climbers' ratings of $1$ unit per year.  The model predicts a climber at a particular time with rating $r_i$ can expect to successfully ascend routes with an identical rating $r_j = r_i$ with probability 50\%.  If that climber's rating improves to $r_i + 1$, the probability of success on the same routes increases to approximately 79\%.

The conventional grade prior weight $b = 0.4$ implies a ratio of 0.4 ratings units per Ewbank grade.  The model predictions actually implied a larger coefficient, but values of $b > 0.4$ caused instability and were considered overly-informative.  Interestingly, the model with $b = 0.4$ was less than 0.1\% more accurate than comparable models with $b = 0$, suggesting that the model is able to recover most of the information in the grades.

The climber prior variance $\sigma_C^2$ and route prior variance $\sigma_R^2$ were chosen as a balance between performance metrics (accuracy and log loss) and numerical stability.  Larger values were generally more accurate, but where $\sigma_C^2$ and $\sigma_R^2$ were greater than 4, extreme values caused instability.

\subsection{Evaluation}

On the Australian dataset with the hyperparameters chosen above, the algorithm detected convergence after 192 iterations (40 seconds using a single thread on an Intel Core i5-8210Y CPU).

Performance metrics were better than baseline, both with and without cross-validation (Table~\ref{tab:metrics}).  The success rate was imbalanced ($\bar{A} = 0.727$) and the model was more accurate for successful ascents (Table~\ref{tab:contingency}).

\begin{figure}[bp]
  \centering
  \input{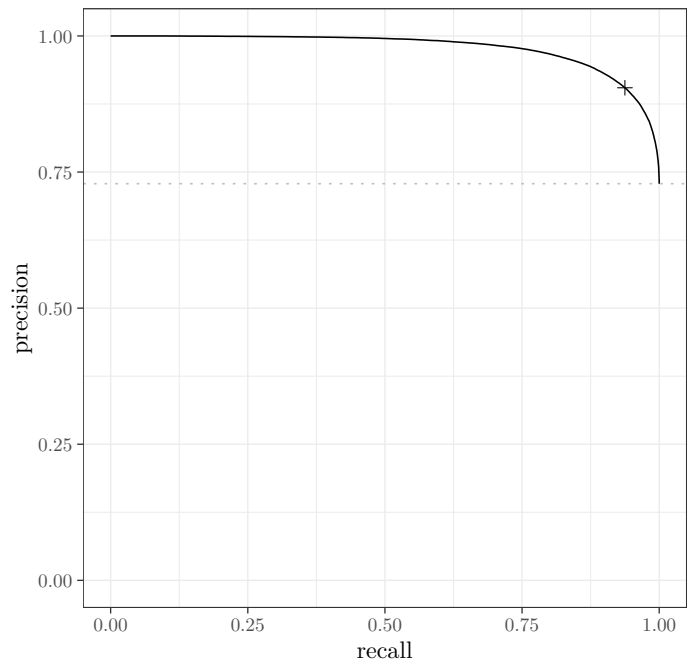}
  \caption{Precision-recall curve for clean ascents, with binary classifier marked (Australian dataset, no cross-validation)}
  \label{fig:precision-recall}
\end{figure}

The precision and recall (without cross-validation) were 90.5\% and 93.7\% respectively.  This favourable position on the precision-recall curve (Figure~\ref{fig:precision-recall}) indicates $P > 0.5$ is an appropriate threshold for the classifier.

\begin{table}[t]
  \caption{Performance metrics for models fitted to the Australian dataset}
  \label{tab:metrics}
  \begin{tabular}{r|ccc}
    \hline
        & Baseline & No cross-validation & Cross-validation \\
    \hline
    Log loss          & 0.585 & 0.270 & 0.373 \\
    Accuracy          & 0.727 & 0.883 & 0.851 \\
    Balanced accuracy & 0.500 & 0.836 & 0.795 \\
    \hline
  \end{tabular}
\end{table}

\begin{table}[t]
  \caption{Binary classifier contingency table for the Australian dataset}
  \label{tab:contingency}
  \begin{tabular}{r|cc}
    \hline
               & Actual success & Actual failure \\
    \hline
    Predicted success & 161,253 & 16,968 \\
    Predicted failure &  10,755 & 47,119 \\
    \hline
  \end{tabular}
\end{table}

\begin{figure}[bp]
  \centering
  \input{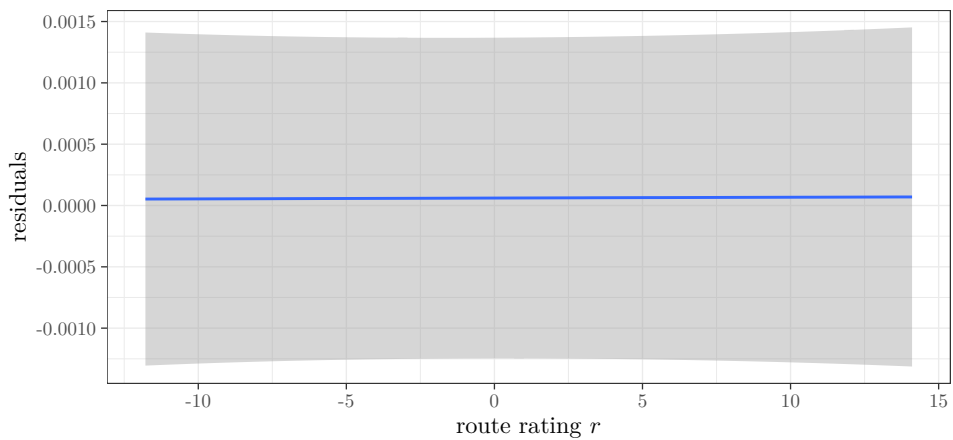}
  \caption{Fitted relationship of residuals to route ratings, with 95\% confidence interval shaded}
  \label{fig:residuals-route-ratings}
\end{figure}

No bias was detected in the relationship between residuals and route ratings, with the average residuals being tightly distributed around zero (Figure~\ref{fig:residuals-route-ratings}).

\begin{figure}[p]
  \centering
  \input{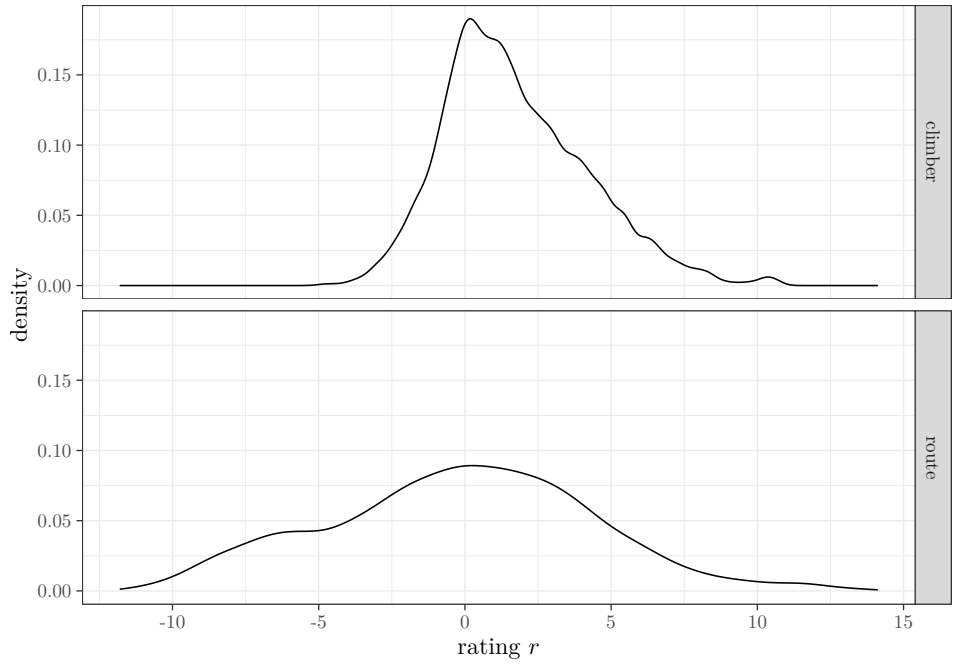}
  \caption{Smoothed distribution of route and climber ratings}
  \label{fig:ratings-density}
\end{figure}

Route and climber ratings estimates ranged between $-12$ and 13, with modes near zero (Figure~\ref{fig:ratings-density}).  The estimated ratings distributions do not resemble their normally-distributed priors: they have non-zero means, wider dispersion, and skewness.

\begin{figure}[p]
  \centering
  \input{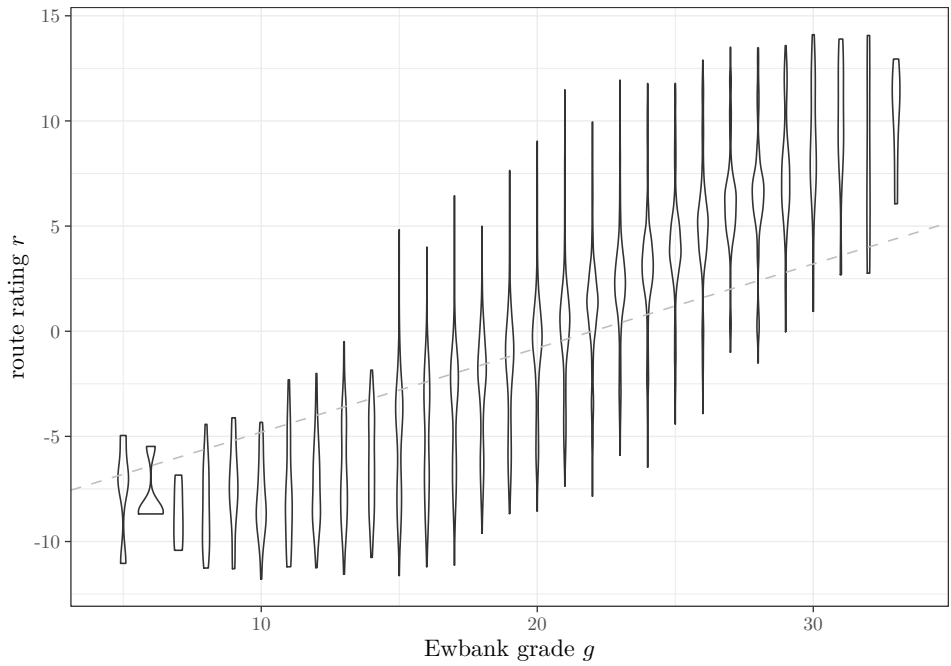}
  \caption{Violin plots of the estimated route ratings at each Ewbank grade; dashed line indicates prior $\mu_i = b(g_i - g_0)$}
  \label{fig:ratings-vs-grades}
\end{figure}

The route ratings estimated by the model have a weak linear relationship ($R^2 = 0.640$) with Ewbank grades.  This is expected to an extent because of the inclusion of the grades in the priors.  Nevertheless, there is a significant overlap in ratings between adjacent grades due to the model's adjustments (Figure~\ref{fig:ratings-vs-grades}).

\section{Discussion}
\label{sec:discussion}

WHR was found to be a pragmatic algorithm for estimating ratings for climbing, given a dynamic Bradley-Terry model.  Its computational performance made it practical to use on a country-scale database of ascents.

The results demonstrate that the model is able to make accurate predictions, and that it is technically well-behaved.  The ratings it learns are correlated with Ewbank grades, but with adjustments that cross multiple grade boundaries.  Interestingly, the model's accuracy is only marginally improved by using Ewbank grades to inform the model's priors.

This paper focuses on ratings and grades for climbing routes.  The algorithm used also generates estimates for climber ratings, which may change over time.  There are opportunities to use this for tracking the progress of climbers for training or ranking purposes.  However, there are also challenges to such applications, due to the reliance on climbers reporting their ascents accurately.

There are opportunities for more sophisticated models.  Another interesting avenue of research would be to use unsupervised learning to identify clusters of climbers and routes with similar traits that affect ascent success.  There are also many known effects that challenge assumptions of the model: holds may break on routes, increasing their difficulty for future ascents; climbers gain knowledge specific to a route, making it easier; specific styles of ascents may affect the difficulty.  These effects could be incorporated into new models.

The author hopes this research can inspire a philosophical shift in the climbing community toward using a data-driven approach to judge the difficulty of routes.

\printbibliography

\end{document}